# Insight into the Properties of the UK Power Consumption Using a Linear Regression and Wavelet Transform Approach

**Samir Avdakovic[1], Alma Ademovic[1] and Amir Nuhanovic[2]**

[1]*EPC Elektroprivreda B&H D.D. Sarajevo, Department for Development, Vilsonovo setaliste 15, 71000 Sarajevo, Bosnia and Herzegovina*
*s.avdakovic@elektroprivreda.ba* and *al.ademovic@elektroprivreda.ba*
[2]*University of Tuzla, Faculty of Electrical Engineering, Department of Power Systems Analysis, Franjevacka 2, 75000 Tuzla, Bosnia and Herzegovina*
*amir.nuhanovic@untz.ba*

**Abstract:** In this paper, the relationship between the Gross Domestic Product (GDP), air temperature variations and power consumption is evaluated using the linear regression and Wavelet Coherence (WTC) approach on a 1971-2011 time series for the United Kingdom (UK). The results based on the linear regression approach indicate that some 66% variability of the UK electricity demand can be explained by the quarterly GDP variations, while only 11% of the quarterly changes of the UK electricity demand are caused by seasonal air temperature variations. WTC however, can detect the period of time when GDP and air temperature significantly correlate with electricity demand and the results of the wavelet correlation at different time scales indicate that a significant correlation is to be found on a long-term basis for GDP and on an annual basis for seasonal air-temperature variations. This approach provides an insight into the properties of the impact of the main factors on power consumption on the basis of which the power system development or operation planning and forecasting the power consumption can be improved.

## 1 INTRODUCTION

For the power system development and operation planning to be efficient, the characteristics of the power consumption should be fully understood. Identification of the main factors affecting power consumption, such as growth and structure of GDP, demographic changes, housing standards, mobility of the population, climate changes, changes in the energy efficiency, habits and customs, etc., is a constant challenge for engineers. Seasonal power consumption variations are usually a result of air-temperature variations [1]-[4], while the long-term analyses show a strong correlation between the economic activity and power consumption. GDP is one of the indicators of the country's economic activity and represents its total output. The impact of the economic activity on energy and power consumption has been a subject of intensive interest for many years and different approaches and analyses can be found in [5]-[15]. In recent years, to allow for a practical analysis of the power consumption, the Wavelet Transform (WT) as a technique for signals and time-series analysis, has been paid a considerable attention by the scientific community. Several works refer to the WT applications for power system load forecasting [16]-[18], while [19] and [20] refer to the power-balancing and wavelet multi-scale analyses of the power-system load variance, respectively. In the context of practical applications of WT in the analysis of power consumption, we refer the reader to [11], [21], [22]. To explore the impact of GDP on power consumption in China, *Jia-Hai et al.* combined the WT and 'Granger causality test' approaches [22], while in the analysis of the Spain power consumption *Gonzalez-Concepcion et al.* apply a discrete WT [21]. *Ozun et al.* use the maximal overlap discrete WT to examine the causality between power consumption and output in the Turkish manufacturing sector. The findings of these studies show that the maximum wavelet correlation between GNP and power consumption can be found at a 5-8 years time-scale [11].

In this paper, the relationship between GDP, seasonal air-temperature variations and power consumption is determined by using real data from the United Kingdom (UK) and a linear regression approach. GDP and air-temperature variations are factors that significantly affect power consumption. Of course, other factors should not be ignored. However, due to the lack of data, the impact of other factors is difficult to analyze. The UK was selected because of the availability of the data on power consumption, GDP and air temperature variations. Other data have been too difficult to obtain. The correlations between these time series at different time-scale spaces are investigated and visualized using the Wavelet Coherence (WTC) approach by means of the Morlet wavelet function [23]-[25]. The results obtained in this paper provide a full insight into the impact of important parameters on power consumption at different time-frequency bands. For power system planners this information can be very useful and the process of operation planning or forecasting can be made more accurate.

## 2 BACKGROUND

Linear regression is a common engineering tool. The related mathematical elaborations will not be dealt with in this paper. On the other hand, WT presents a relatively young mathematical technique and detailed mathematical explanations and practical application in various areas of science can be found in many books and papers. The selection of an appropriate wavelet function or "mother wavelet" and using shifted and dilated versions of the same wavelet present the basic principle of the wavelet approach, allowing analysis to be performed at different frequency/period bands with different resolutions [26]-[30]. As said above, to illustrate the impact of GDP and air-temperature variations on power consumption at different time-scales, we use the WTC approach [23]. This approach is often used in determining the relationship between two time series [24]-[25].

We follow [23]-[25], [31]-[35] and define the basic mathematical expressions of the methodology used in this paper. Morlet wavelet function $\psi(t)$ used in this study is defined as:

$$\psi(t) = \pi^{-1/4} e^{i\omega_0 t} e^{-t^2/2}, \quad (1)$$

where: value $\pi^{-1/4}$ – is a normalization factor, $t$ – is the dimensionless time parameter, and $\omega_0$ – is the dimensionless frequency parameter. Using the value of $\omega_0 = 6$, the Morlet wavelet provides a great balance between time and frequency localization. For this central frequency of the wavelet function, the Fourier frequency period is almost equal to the scale ($f \approx 1/s$). Also, the Morlet wavelet is non-orthogonal and complex [31]-[35].

Continuous WT (CWT) of time series $x_n$, ($n = 1, 2, 3, \dots, N$) is with wavelet function $\psi(t)$ and equal time intervals $\delta_t$ defined as [23]-[25]:

$$W_m(s) = \frac{\delta_t}{\sqrt{s}} \sum_{n=0}^{N-1} x_n \psi * \left[\frac{(n-m)\delta_t}{s}\right], \quad (2)$$

where: $m = (0, 1, \dots, N-1)$, $*$ – denotes the conjugate complex value, $N$ – is the number of points in the evaluated time series and $\psi(t)$ – is the wavelet function at scale $s$ translated in time by $m$ [25].

The wavelet power spectrum ($|W_n(s)|^2$) presents the squared absolute value of the wavelet coefficients (or squared amplitude), while the cross-wavelet transform of two time series, $x_n$ and $y_n$, is defined as $W_n^{xy} = W_n^x W_n^{y*}$, where $W_n^x$ and $W_n^y$ are WT of the $x$ and $y$ time series, respectively. The value $|W_n^{xy}|$ represents the cross-wavelet power. At each defined scale, the cross-wavelet power of two time series suggests the local covariance between the analyzed data sets and provides an insight into their power similarity [23]-[25]. However, the cross-wavelet power is the product of two non-normalized wavelet spectra and can lead to erroneous conclusions [34]. In this way it is possible to identify the significant cross-wavelet spectrum between two time series, although there is no significant correlation between them. WTC overcomes this problem by normalizing the single-wavelet power spectrum [34]. WTC between two time series, $x_n$ and $y_n$, is defined as [34]:

$$WTC = \frac{|W_n^{xy}(s)|}{[W_n^x(s) W_n^{y*}(s)]^{0.5}}, \quad (3)$$

where the notation corresponds to that in the expression for the cross-wavelet transform.

In this study, WTC is used as a measure of intensity of the covariance of the two time series (GDP or air temperature and power consumption) in the time-frequency space or the local correlation between the time series in the time-frequency space. WTC ranges from 0 to 1 and it can be interpreted as a correlation coefficient; the closer the value is to 1 the more correlated are the two series [34]. This approach has proven to be a very useful tool in the analysis of time-series. The phase-difference between the two time series is defined as an argument of the smoothed real and imaginary fragments of the cross-spectrum [33] and it provides information on the delays of the oscillations between the two time series. More details about the approach used in this paper can be found in [23]-[25] and [31]-[35]. The software used in this paper is available in [36].

## 3 RESULTS AND DISCUSSIONS

The main results and findings of our analysis performed for the selected data sets are presented in this section. To analyze and illustrate the impact of GDP and air-temperature variations on power consumption, data from the UK are used. The data of the average seasonal air-temperature variations, quarterly GDP and quarterly power consumption in the UK for the period 1971-2011 are presented in Fig. 1 and are taken over from [37], [38] and [39], respectively. The data of the air-temperature variations present the average value of seasonal temperature measurements at various locations throughout England. This region is wide enough to justify the use of these data for the total power consumption and reflects its seasonal air-temperature variations.

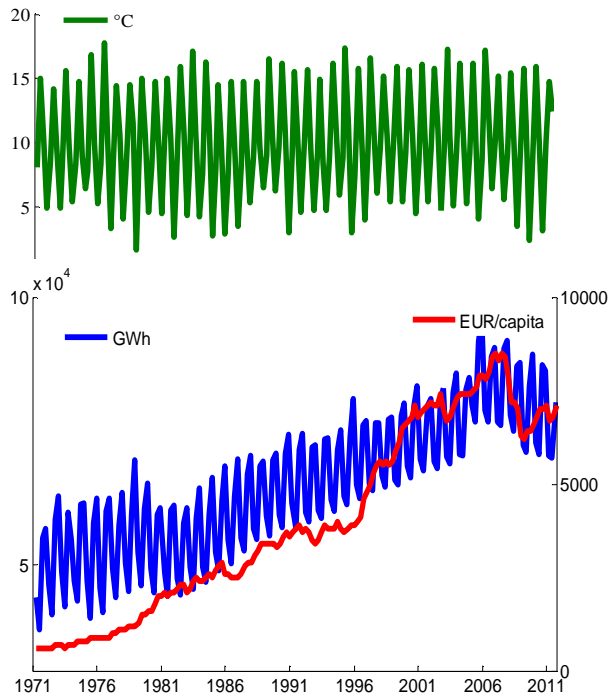

Figure 1. UK average seasonal air-temperature variations and quarterly values of GDP and power consumption for the period 1971-2011

In general, the UK presents one of the economically most developed countries of the world which in recent years has recorded a decrease in the values of GDP and power consumption (Fig. 1) as a results to the global economic crisis. However, from Fig. 1 a slight decrease in power consumption at the end of the seventies can also be identified. This phenomenon could be related to the energy crisis in this period.

The correlation between the power consumption and GDP for the time series from Fig. 1, determined by using the linear regression approach, is presented in Fig. 2, while the correlation between the power consumption and seasonal air-temperature variations, also for the time series from Fig. 1 is presented in Fig. 3.

The relation between GDP and the power consumption using the linear regression is given by equation $y = ax+b$, where: $y$ – present the power consumption (GWh), $x$ – GDP (EUR/capita) and $a$ and $b$ – are the respective regression coefficients. The equations describing the relations between these two variables are presented in Fig. 2. The value of the Pearson correlation for this model is 0.8157 and it can be observed that dissipation in the power consumption in the UK is relatively small. The R-square ($R^2$) or the coefficient of determination is a coefficient used for evaluation of the representativeness of the regression model. It is based on the analysis of the respective variance and for the UK power consumption and GDP time series its value is 0.665 (Fig. 2).

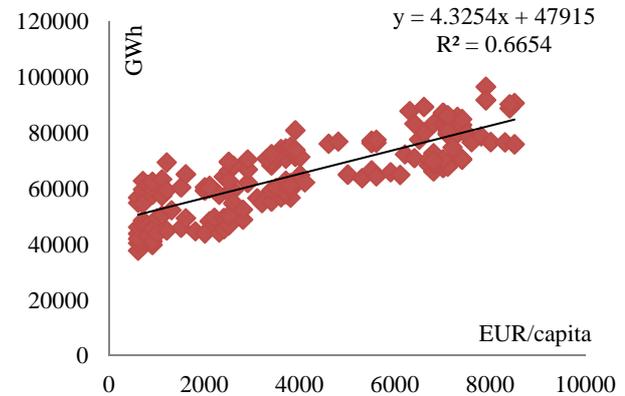

Figure 2. Correlations between the UK power consumption and GDP

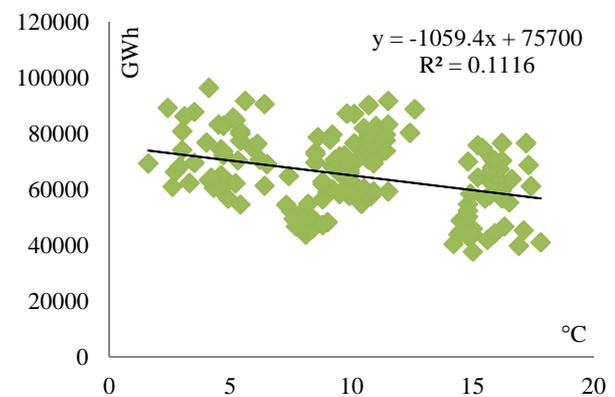

Figure 3. Correlations between the UK power consumption and seasonal air-temperature variations

In other words, 66.5% of the UK power consumption variations can be explained by the quarterly GDP variations. For the UK time series plotted in Fig. 2 as the quarterly power consumption against the quarterly GDP, the fitted curve shows the GDP sensitivity on the power consumption at about 4.33 GWh/(EUR/capita), with a relatively high coefficient of determination, approximately 66%. Given that for the observed time series the UK quarterly GDP variations can be several hundred EUR/capita, the quarterly change in the value of the power consumption due to variations in GDP might be significant. On the other hand, analyzing the correlation between the power consumption and seasonal air-temperature variations, it can be noted that it has a negative trend and a low value of $R^2$ (Fig. 3). The fitted curve shows the seasonal air-temperature sensitivity of the power consumption at about 1060 GWh/°C. Also, it is evident that there are three groups of data gathered around the seasons: winter, summer, and spring and autumn. It is clear that there is a significant positive correlation between the data for spring and autumn. Generally, without going into a detailed analysis of subgroups, analyzing the entire time series and all subgroups within them, a negative correlation is obtained. However, it is obvious that the linear regression is an approach too simplistic for this type of applications.

To examine and illustrate the relationship between two time series over time, the WTC approach is quite an efficient tool. The WTC and phase analysis between the GDP and the power consumption are presented in Fig. 4, while WTC and phase analysis between seasonal air-temperature variations and the power consumption are presented in Fig. 5.

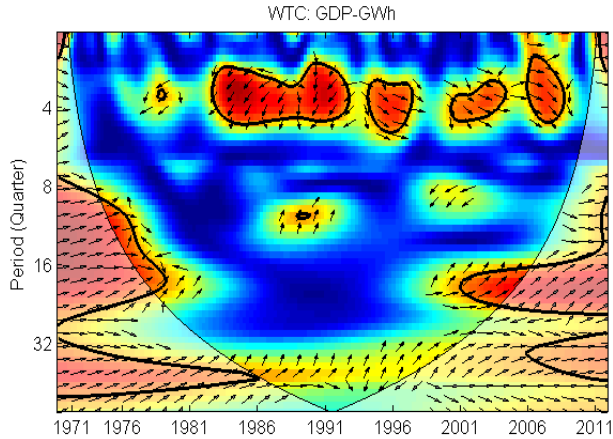

Figure 4. WTC and phase difference between GDP and the power consumption

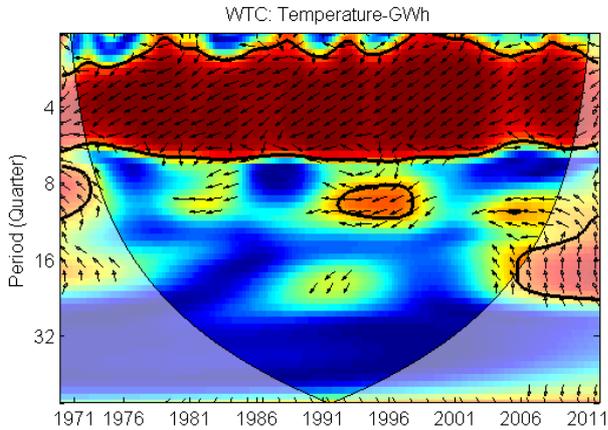

Figure 5. WTC and phase difference between seasonal air-temperature variations and the power consumption

The phase relationship is given in the form of arrows (the in-phase pointing to the right and the anti-phase pointing to the left), while the colour code indicates different coherency ranges, ranging from blue to red, adding to the significance of the regions [24], [25]. Significant wavelet coherency between GDP and the power consumption was found for several periods throughout the observed time horizon. For the four-quarter or one-year period band, the GDP and the power consumption show a significant coherence for the period 1982-1992, with anti-phase correlations. With a phase difference in the quadrant ($-\pi,-\pi/2$) [25], the power consumption is lagging GDP for this period.

Also, for the same quarter periods, significant coherence can be identified around 1995, 2002 and 2007, and all of these correlations are in-phase, suggesting a positive correlation. The GDP index and power consumption also show a significant coherence during the periods of 1971-1978 and 2000-2011 with a periodicity between 8–30 quarters and the correlations in-phase, which again suggests a positive correlation. Finally, for the period band over 32 quarters, the UK GDP index and the power consumption show a significant coherence almost over the entire observed time horizon, with in-phase correlations indicating strong relations between these two time series. Also, one can notice that the arrows pointing right and slightly up suggest that power consumption is lagging compared to the GDP index over the observed time horizon. This shows the high dependence of power consumption on the GDP index, especially for the higher period bands (of about 8 years). The significant regions presented in Fig. 4 provide a clearer insight into the relationship between these two time series.

Looking at Fig. 5 which shows the WTC and phase difference between the time series of the seasonal air-temperature variations and the power consumption, several regions with significant correlation can be identified. The most significant and out of phase region over the entire observed time horizon is for the period in the 2-6 quarter band. This clearly indicates a strong seasonal relationship between these two time series. In other words, the lower temperature during the winter period will result in higher power consumption, while the higher temperatures during the summer days will decrease it. The second important region, out of phase as well, is identified for the period of 1992-1997 in the 8-10 quarter band. The reason for this is the fact that in this period 'peaks' in the time series can be identified (Fig. 1). The third significant region is identified in the period of 2005-2011 for the 16-20 quarter band. The phases in this region are oriented at $\pi/2$ which indicates that these two time series in this time interval are independent. This coincides with the outbreak of the global economic crisis and the beginning of the decrease in the UK power consumption.

## 4 CONCLUSION

In this paper, a linear regression and WTC approach are applied to analyze the impact of GDP and seasonal temperature variations on the UK power consumption. To our knowledge, this paper is the first one to research the relationship between GDP, air temperatures and the power consumption using WTC. For selected time series, the results of the linear regression approach shows the GDP sensitivity to the power consumption of some 4.33 GWh/(EUR/capita), with a relatively high determination coefficient, of approximately 66%. The seasonal air temperature sensitivity to the power consumption is of some 1060 GWh/°C. However, this approach can provide only a global observation of the impact of these parameters on the power consumption; an ultimate conclusion would require other approaches.

WTC and phase analysis present a very useful tool to analyse the relationship between the two time series at different time-scales. From the analysis performed in this research, which is related to the UK power consumption, the following conclusions can be drawn:

- The seasonal and annual UK power consumption are highly connected with the air temperature variations. The results of the data analysis also show that in the past there was a period on the 3-5 quarter band when GDP, too, had an impact on the power consumption.

- For the period of the 4-8 quarters (1-2 years), the results show that there is no relationship between the UK GDP and the power consumption, while positive correlations are identified for the period of 16 quarters (4 years) and the intervals of 1971-1978 and 2000-2011. A significant positive correlation is identified for the entire time interval for a period of about 32 quarters (8 years), indicating that the economic activity of a country on a longer time-scale significantly affects the rise/fall in power consumption.

As seen from the previous conclusions, forecasting the GDP and seasonal air-temperature index trends should be taken into account in mid and long-term forecasting and power-system planning. This is the task to be dealt with in our future research.

## REFERENCES

[1] Pilli-Sihvola, K., Aatola, P., Ollikainen, M., and Tuomenvirta, H., Climate change and electricity consumption-witnessing increasing or decreasing use and costs?, *Energy Policy*, Vol. 38, 2010, pp. 2409-2419.

[2] Parkpoom, S. J., and Harrison, G. P., Analyzing the impact of climate change on future electricity demand in Thailand, *IEEE Trans on Power Systems*, Vol. 23, 2008, pp. 1441-1448

[3] Valor, E., Meneu, V., and Caselles, V., Daily air temperature and electricity load in Spain, *Journal of Applied Meteorology*, Vol. 40, 2001, pp. 1413-1421.

[4] Henley, A., and Peirson, J., Non-linearities in electricity demand and temperature: Parametric versus non-parametric methods, *Oxford Bulletin of Economics and Statistics*, Vol. 59, 1997, pp. 149-162.

[5] Georgantopoulos, A. G., and Tsamis, A. D., The relationship between energy consumption and GDP: a causality analysis on Balkan countries, *Eur Journal of Scientific Research*, Vol. 61, 2011, pp. 372-380.

[6] Kankala, M., Akpinar, A., Komurcu, M. I., and Ozsahin, T. S., Modeling and forecasting of Turkey's energy consumption using socio-economic and demographic variables, *Applied Energy*, Vol. 88, 2011, pp. 1927-1939.

[7] Bekhet, H. A., and Othman, N. S., Causality analysis among electricity consumption, consumer expenditure, gross domestic product (GDP) and foreign direct investment (FDI): Case study of Malaysia, *Journal of Economics and International Finance*, Vol. 3, 2011, pp. 228-235.

[8] Ighodaro, C. A. U., Co-integration and causality relationship between energy consumption and economic growth: further empirical evidence for Nigeria, *Journal of Business Economics and Management*, Vol. 11, 2010, pp. 97-111.

[9] Huang, B. N., Hwang, M. J., and Yang, C.W., Causal relationship between energy consumption and GDP growth revisited: A dynamic panel data approach, *Ecological economics*, Vol. 67, 2008, pp. 41-54.

[10] Hou, Q., The relationship between energy consumption growths and economic growth in China, *Inter Journal of Economics and Finance*, Vol. 1, 2009, pp. 232-237.

[11] Ozun, A., and Cifter, A., Multi-scale causality between energy consumption and GNP in emerging markets: evidence from Turkey, *Investment Management and Financial Innovations*, Vol. 4, 2007, pp. 60-70.

[12] Chen, S. T., Kuo, H. I., and Chen, C. C., The relationship between GDP and electricity consumption in 10 Asian countries, *Energy Policy*, Vol. 35, 2007, pp. 2611-2621.

[13] Fung, W. Y., Lam, K. S., Hung, W. T., Pang, S. W., and Lee, Y. L., Impact of urban temperature on energy consumption of Hong Kong, *Energy*, Vol. 31, 2006, pp. 2623-2637.

[14] Chima, C. M., and Freed, R., Empirical study of the relationship between energy consumption and gross domestic product in the U.S.A', *Inter Business&Economics Research Journal*, Vol. 4, 2005, pp.101-112.

[15] Soytas, U., and Sari, R., Energy consumption and GDP: causality relationship in G-7countries and emerging markets, *Energy Economics*, Vol. 25, 2003, pp. 33-37.

[16] Zhang, Q., and Liu, T., Research on mid-long term load forecasting base on wavelet neural network, in Proc. of the Second Int. Conf. on Computer Engineering and Applications, 2010, pp. 217-20.

[17] Khoa, T. Q. D., Phuong, L. M., Binh, P. T. T., and Lien N. T. H., Application of wavelet and neural network to long-term load forecasting, in Proc. *of the Int. Conf. on Power System Technology-POWERCON*, 2004, pp. 840-44.

[18] Senjyu, T., Tamaki, Y., Takara, H. and Uezato, K., Next day load curve forecasting using wavelet analysis with neural network, Electric Power Components and Systems, Vol. 30, 2002, pp. 1167-1178.

[19] Frunt, J., Kling, W. L., and Ribeiro, P. F., Wavelet decomposition for power balancing analysis, *IEEE Trans. on Power Delivery*, Vol. 26, 2011, pp. 1608-1614.

[20] Avdakovic, S., Nuhanovic, A., Kusljugic, M., Becirovic, E., and Turkovic, E., Wavelet multiscale analyses of a power system load variance, *Turkish Journal of Electrical Eng & Comp Sci*, Accepted for publication, 2012, DOI: 10.3906/elk-1109-47, 2012.

[21] Gonzalez-Concepcion, C., Gil-Farina, M. C., and Gabino, C. P., Wavelets modelling of water, energy and electricity consumption in Spain, *WSEAS Trans. on Mathematics*, Vol. 9, 2010, pp. 509-518.

[22] Jia-Hai, Y., Chang-Hong, Z., & Min-Peng, X., Frequency granger causality test in cointegration system by wavelet analysis, *in Proc. of the 10th WSEAS Int. Con. on Applied Mathematics*, 2006, pp. 208-13.

[23] Grinsted, A., Moore, J. C., and Jevrejeva, S., Application of the cross wavelet transform and wavelet coherence to geophysical time series, *Nonlinear Processes in Geophysics*, Vol. 11, 2004, pp. 561-566.

[24] Sen, A. K., Zheng, J., and Huang, Z., Dynamics of cycle-to-cycle variations in a natural gas direct-injection spark-ignition engine, *Applied Energy*, Vol. 88, 2011, pp. 2324-34.

[25] Aguiar-Conraria, L., Azevedo, N., and Soares M. J., Using wavelets to decompose the time–frequency effects of monetary policy, *Physica A*, Vol. 387, 2008, pp. 2863–2878.

[26] Avdakovic, S., Nuhanovic, A., Kusljugic, M., and Music, M., Wavelet transform applications in power system dynamics, *Electric Power Systems Research*, Vol. 83, 2012, pp. 237-245.

[27] He, H. and Starzyk, J. A., A self-organizing learning array system for power quality classification based on wavelet transform, *IEEE Trans. on Power Delivery*, Vol. 21, 2006, pp. 286–295.

[28] Mallat, S., A wavelet tour of signal processing, San Diego, CA: Academic Press, Inc, 1998.

[29] Vetterli, M., and Kovacevic, J., Wavelets and subband coding, New York: Prentice-Hall, Inc., 1995.

[30] Daubechies, I., Ten lectures on wavelets, Philadelphia: Society for Industrial and Applied Mathematics, 1992.

[31] B. Cazelles, M. Chavez, G.C. de Magny, J.F. Guegan, S. Hales, Time-dependent spectral analysis of epidemiological time-series with wavelets, *Journal of the Royal Society Interface*, Vol. 4, 2007, pp. 625–636.
[32] V.W. Keener, G.W. Feyereisen, U. Lall, J.W. Jones, D.D. Bosch, R. Lowrance, El-Niño/Southern Oscillation (ENSO) influences on monthly NO3 load and concentration, stream flow and precipitation in the Little River Watershed, Tifton, Georgia (GA), *Journal of Hydrology*, Vol. 381, 2010, pp. 352–363.
[33] A. C. Furon, C. Wagner-Riddle*, C. Ryan Smith, J. S. Warland, Wavelet analysis of wintertime and spring thaw $CO_2$ and $N_2O$ fluxes from agricultural fields, *Agricultural and forest meteorology*, Vol. 148, 2008, pp. 1305-1317.
[34] I. P. Holman, M. Rivas-Casado, J. P. Bloomfield, J. J. Gurdak, Identifying non-stationary groundwater level response to North Atlantic ocean-atmosphere teleconnection patterns using wavelet coherence, *Hydrogeology Journal*, Vol. 19, 2011, pp. 1269-1278.
[35] J. Bigot, M. Longcamp, F. Dal Maso, D. Amarantini, A new statistical test based on the wavelet cross-spectrum to detect time–frequency dependence between non-stationary signals: Application to the analysis of cortico-muscular interactions, *NeuroImage*, Vol. 55, 2011, pp. 1504–1518.
[36] Av.: http://www.pol.ac.uk/home/research/waveletcoherence/
[37] Parker, D.E., T.P. Legg, and C.K. Folland. 1992. A new daily Central England Temperature Series, 1772-1991. Int. J. Clim., Vol 12, pp 317-342, Available: www.metoffice.gov.uk/hadobs
[38] Available: http://epp.eurostat.ec.europa.eu/
[39] Available: http://www.nationalgrid.com/